\documentclass[a4paper]{jpconf}
\usepackage{graphicx}
\usepackage{cite}
\begin{document}

\title{On the efficiency of the evaluation of the primary cosmic ray composition using lateral distributions of air shower electromagnetic component}

\author{Roman Raikin, Tatyana Serebryakova, Nikolay  Volkov and Anatoly~Lagutin}

\address{Altai State University. 61 Lenin Ave., Barnaul, 656049, Russia}

\ead{raikin@theory.asu.ru}

\begin{abstract}
New analysis of lateral distributions (LD) of electrons, muons and charged particles in extensive air showers generated by $10^{15}-10^{19}$~eV cosmic rays has been made with respect to the scaling formalism for LD and air shower universality paradigm. It is found that lateral distributions of electrons and muons, when considered separately, are well described by one-parametric scaling functions, while for the charged particles mixture distributions the scaling description is not accurate in radial distance range where electrons and muons give a comparable contribution to the local particle densities. The use of scaling formalism enables enhancing the electron/muon separation capabilities and, when combined with theoretically motivated air shower universality concept, provides robust mass composition estimates. The proposed approach could be implemented for the present and future (multi-) hybrid air shower observations, especially in realizing the potential of Auger and Telescope Array observatories upgrades, as well as for re-analysis and cross-calibration of the data collected from different air shower arrays in a broad primary energy range.
\end{abstract}

\section{Introduction}
The evaluation of primary cosmic ray mass composition is at present stage the crucial issue for the interpretation of the experimental data in terms of sources and propagation models~\cite{Aloisio:2014,Aloisio:2017}. The mass composition is inferred from the extensive air shower (EAS) observations by comparisons with simulations results, which rely on hadronic interaction models fundamentally uncertain in the relevant energy range. Numerous methods and techniques are implemented, including analysis of mean values, fluctuations, correlations or even particular features of distributions of different observables characterized both longitudinal and spatial shower development (depth of maximum, muon production depth, total number of electrons and muons at the observation level and their local densities at various distances from the shower axis, particles arrival time profiles, spatial distribution of radio emission, Cherenkov light)~(see \cite{KampertUnger,Haungs:2015,ABBASI201549,YUSHKOV201629,Aab:2016288,yushkov:2017,Horandel:2017,Chiavassa:2018,Hanlon:2018,PhysRevD.98.103002}).

Large efforts have been made recently in both gaining experimental data with increased resolution in detection of various EAS components and developing improved methods for physical interpretation of the data along with evolving hadronic interaction models after the LHC results. Nevertheless, the composition results remain ambiguous in the entire energy range available for EAS studies. 

The discrepancies between estimates of composition derived by various methods from the data of different experiments are apparently caused by a complex of instrumental and methodical systematic biases of different nature as well as by strong model dependence of observables, mostly in case of muon component characteristics, the so-called {\em muon problem} (for the up-to-date state one can refer to~\cite{MuonExcessPhysRevLett.117.192001,
MuonExcessPhysRevD.98.022002}), disadvantages in taking into account meteorological effects etc.

In this paper we report on the further development of the approach based on the air shower universality and scaling formalism for lateral distribution (LD) that exhibits the potential in improved primary composition evaluation from the 100\% duty cycle surface detectors data~{\cite{Raikin:2008zz,Raikin:2009zz,Raikin:2011dga,Raikin2011,Lagutin13,Lagutin:2013jpcs,Raikin:2016apd,Raikin2017}. 

The paper is organized as follows. In Section 2 the universality in air shower development and scaling formalism for LD are explained. The Monte-Carlo simulations and results of analysis are presented in Section 3.  Section 4 contains brief discussion and concluding remarks.

\section{Air shower universality and scaling formalism}

The concept of air shower universality is inspired by the classical results of the electromagnetic cascade theory and in a broad sense expressed in the similarity of the spectra of low-energy secondary particles with respect to the cascade development stage. As a consequence, the relations between various shower characteristics represented using the cascade age parameter demonstrate to one degree or another the invariance with respect to the properties of a primary particle. Different aspects of the universality (phenomenology, possible generalisations, deeper understanding of the origin and limitations, sensitivity to the hadronic interaction model,
applicability under conditions of real experiments) are subjected to remarkable interest~\cite{Giller:2005,Nerling:2006,Capdevielle:2005,Gora:2006,Apel:2008,Lipari:2009,Lafebre:2009,Mattews:2010,Yushkov:2010,Ivanov:2011,Tapia:2013,Capdevielle:2012,Giller:2015,Dey:2016shg,Bridgeman:2017rcv, Bartoli:2017,Giller:2018}. 

In our papers~\cite{Lagutin:2002tv,Raikin:2008zz,Raikin:2009zz,Raikin:2011dga,Raikin2011,Lagutin13,Lagutin:2013jpcs,Raikin:2016apd,Raikin2017} it was shown that on the basis of scaling formalism for lateral distribution of electrons in air showers one can observe the generalized manifestation of the universality very promising for model-independent EAS reconstruction in a wide primary energy range.

The basic idea of the one-valued relation between the shape of electron LDF and shower age dates back to the classical Nishimura-Kamata-Greizen (NKG) function~\cite{NKG1,NKG2}:
\begin{equation}
\rho(r;E,s)=N(E,s)\frac{C}{r_0^2}
\left(\frac{r}{r_0}\right)^{s-2}\left(1+\frac{r}{r_0}\right)^{s-4.5}.
\label{NKG}
\end{equation}
Here $\rho(r;E,s)$ is local particle density at radial distance
$r$ from the core position in shower with primary energy $E$ and the longitudinal age parameter~$s=3X/(X+2X_{\rm max})$, where $X$ is the observation depth in the atmosphere, $X_{\rm max}$ is the depth of shower maximum, $N(E,s)$~-- total number of particles at the observation depth (shower size), $r_0$~-- constant shower scale radius, which does not depend on primary particle type and energy (originally~-- the Moliere unit $r_M$).

The NKG function is insufficient to describe electron LDFs accurately at large distances~(see e.g.~\cite{Lagutin:1979}). This led to various modifications of the NKG form, such as changing the values of exponents, introducing different constant or variable scale factors, lateral ($s_\perp$) or local ($s(r)$) age parameters and also generalizations of the function by using a third power-law term. A comprehensive review is beyond the scope of this paper (some discussions can be found in~\cite{Capdevielle:2012,Dey:2016shg,Tapia:2013,Bartoli:2017,Apel:2006ld}).

According to the scaling approach introduced in~\cite{Lagutin:1997,Lagutin:2002tv} the lateral distribution of electrons in both gamma- and hadron-induced air showers can be properly described by the scale-invariant function:
\begin{equation}
\rho(r;E,X)=\frac{N(E,X)}{R_0^2(E,X)}\,F\left(\displaystyle\frac{r}{R_0(E,X)}\right).
\label{scaling1}
\end{equation}
The principal difference of function (\ref{scaling1}) from the original NKG form~(\ref{NKG}) and its modifications mentioned above is that the scale factor $R_0$ is variable, while function $F(x)$ does not depend explicitly on primary particle properties and observation depth. Consequently, variations of the  shape of lateral distribution are totally described by rescaling the lateral distance.

According to our calculations based on different methods and codes~\cite{Lagutin:1997,Lagutin:2001nc,Lagutin:2002tv,Lagutin13,Lagutin:2013jpcs}, the scaling approach is valid for the lateral distribution of electrons over the interval of scaling variable $x=r/R_0=(0.05-20)$, corresponding to the region of radial distances from $r\sim 10$~m to $r\sim (2-4)$~km depending on shower age. The energy range where scaling formalism can be used extends up to $10^{22}$~eV for electromagnetic showers taking into account Landau-Pomeranchuk-Migdal effect and interactions with geomagnetic field~\cite{Lagutin13,Lagutin:2013jpcs}. For nuclei-initiated extensive air showers the energy region verified by simulations is $(10^{15}-10^{19})$~eV. It is also worth mentioning that for lateral distribution of electrons the scale factor $R_0$ is equal to the root mean square radius of electron component $R_{\rm ms}$, which is defined in a standard way as
\begin{equation}
R_{\rm{ms}}(E,X)=\left(\displaystyle\frac{2\pi}{N(E,X)}\int_0^\infty r^2 \rho(r;E,X)rdr\right)^{1/2},
\end{equation}
and the following expression suggested in~\cite{Lagutin:2001nc} for $F(x)$ could be used
\begin{equation}
F(x)=Cx^{-\alpha}(1+x)^{-(\beta-\alpha)}(1+(x/10)^\gamma)^{-\delta},
\label{lrf}
\end{equation}
with the set of constant parameters $C=0.28$, $\alpha=1.2$, $\beta=4.53$, $\gamma=2.0$, $\delta=0.6$.

The relation between scale factor $R_0$ and cascade age parameter $s$ can be well reproduced by the expression $R_0=\eta f(s)$, where $f(s)$~-- the universal function, $\eta(E,X,\rho(X))$~-– correction factor, describing the influence of atmospheric density profile. Thus, for constant atmospheric conditions at fixed observation depth we have one-valued mapping between lateral distribution shape and shower age $s$ as well as depth of maximum $X_{\rm max}$, that could be used as a theoretical basis for primary composition deduction from surface detectors data as well as from hybrid measurements.

The function of the form~(\ref{lrf}) was implemented by several groups for the description of LD of electrons, muons and charged particles in different energy and radial distance ranges~(see e.g.~\cite{ANTONI2001245,Cotzomi:2007twa,Ivanov2007,Kalmykov2007,Fomin2008,KNURENKO2008201,Apel:2006ld,APEL2010202,Sabourov:2011,refId0,APEL201725,Bartoli:2017}). 
Nevertheless, the use of the shape of the lateral distribution to infer the mass composition is not straightforward due to the impossibility of high-precision measurements of the lateral distribution in a wide range of radial distances, the insufficient capabilities for effective discrimination of the electron contribution to the local densities of charged particles far from the shower core and also the limitations of the scaling approach itself.

\section{Simulations and results}

In this paper we present the results of the analysis of Monte-Carlo simulations of EAS initiated by protons and iron nuclei in the energy range of $10^{15}-10^{19}$~eV performed using CORSIKA v.7.4100~\cite{CORSIKA:1998} with EPOS LHC v.3400 and QGSJet-II-04 (FLUKA 2011.2c.2) hadronic interaction models. For the first time electron, muon and charged particles lateral distributions were considered separately in the framework of scaling and universality approach described above. In order to get reliable data on local particle densities at very large distances from the shower core the thinning level and particle weight limit were set as $\varepsilon_{\rm th}= 10^{-8}$ and $\omega=10^2$ respectively.

Radial scale factors were evaluated for averaged and individual showers by fitting simulated LDs. Note, that to obtain the unbiased estimates of $R_0$ an iterative procedure was implemented for discrimination of data at distances where the scaling formalism is not valid and also for using fixed distances ranges with respect to the scaling variable $x=r/R_0$ for showers of a specific energy and primary particle type. For $F(x)$, different expressions of function~$F(x)$ were used. We have checked out function~(\ref{lrf}) with a refined set of parameters $\alpha,~\beta,~\gamma,~\delta$ and other representations, including polynomial approximation for $x^2F(x)$ giving a considerably better overall fit of simulated data. It was found that the resulting $R_0$ values demonstrate only a weak dependence on the explicit form of $F(x)$ choosing for fitting.

\begin{figure}[t]
\begin{minipage}{18pc}
{\footnotesize a)}\includegraphics[width=18pc]{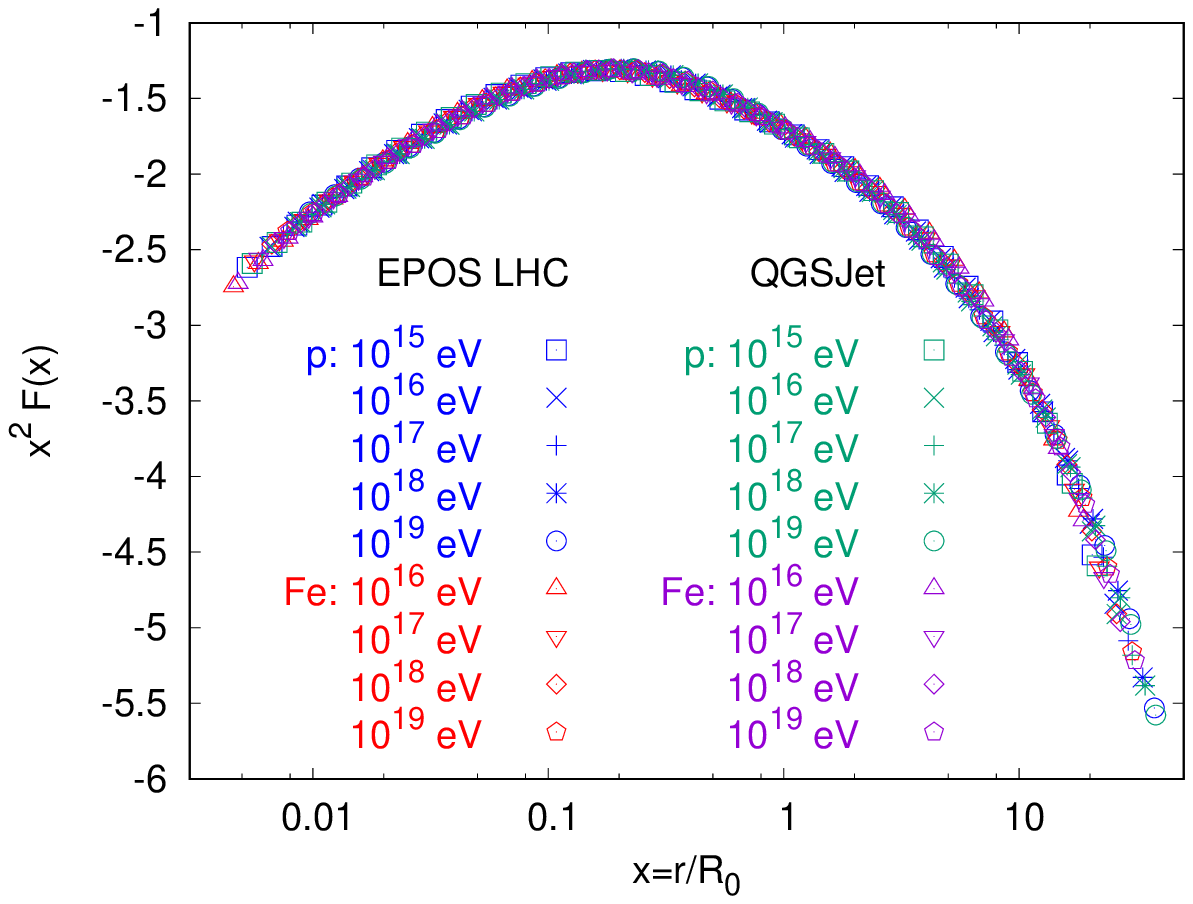}
\end{minipage}\hspace{2pc}%
\begin{minipage}{18pc}
{\footnotesize b)}\includegraphics[width=18pc]{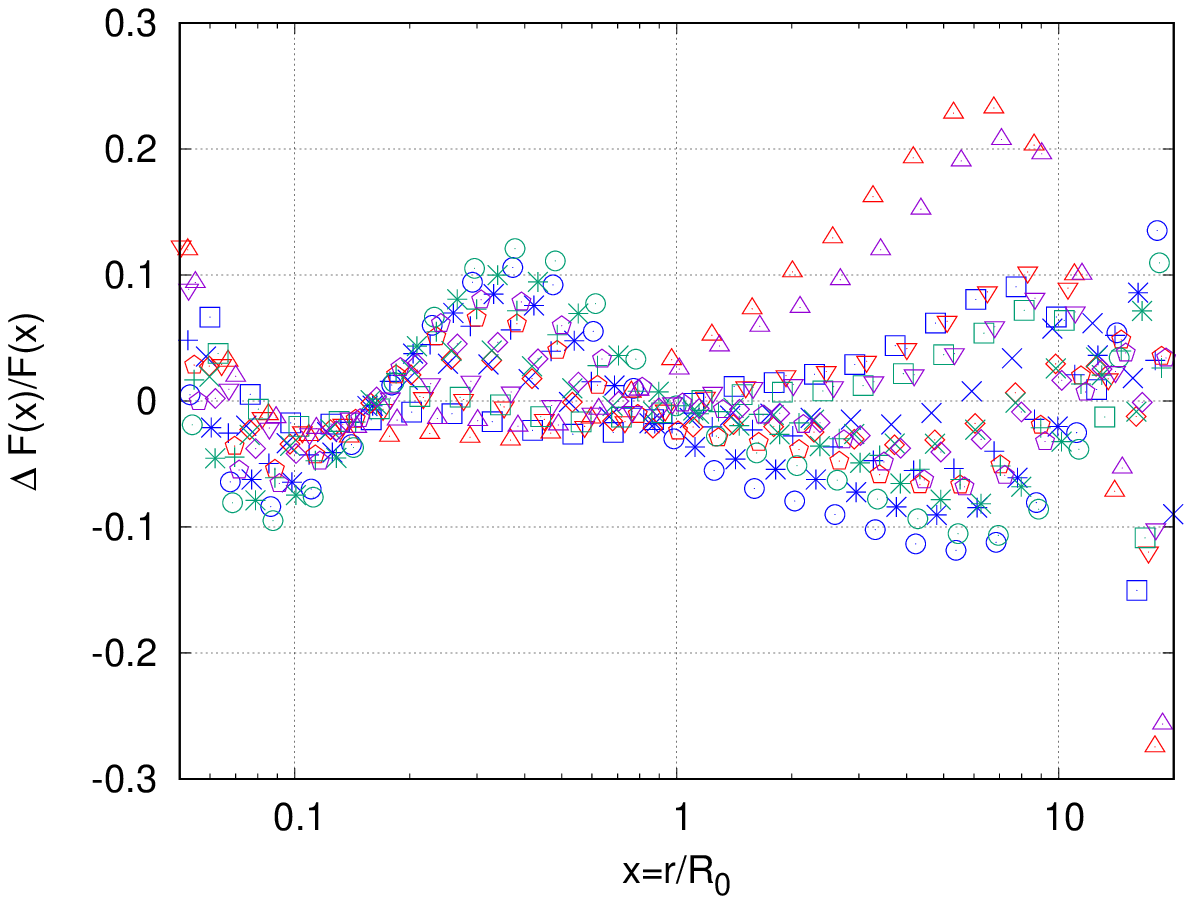}
\end{minipage} 
\caption{\label{f1}Invariant part (a) and relative uncertainties (b) of the scaling description of average lateral distribution of electrons at sea level for vertical simulated EAS. Results obtained for proton and iron primaries in the energy range of $\left(10^{15}-10^{19}\right)$~eV using EPOS LHC and QGSJet hadronic interaction models are shown together by coloured symbols described in the Figure. See text for details.}
\end{figure}

\begin{figure}[t]
\begin{minipage}{18pc}
{\footnotesize a)}\includegraphics[width=18pc]{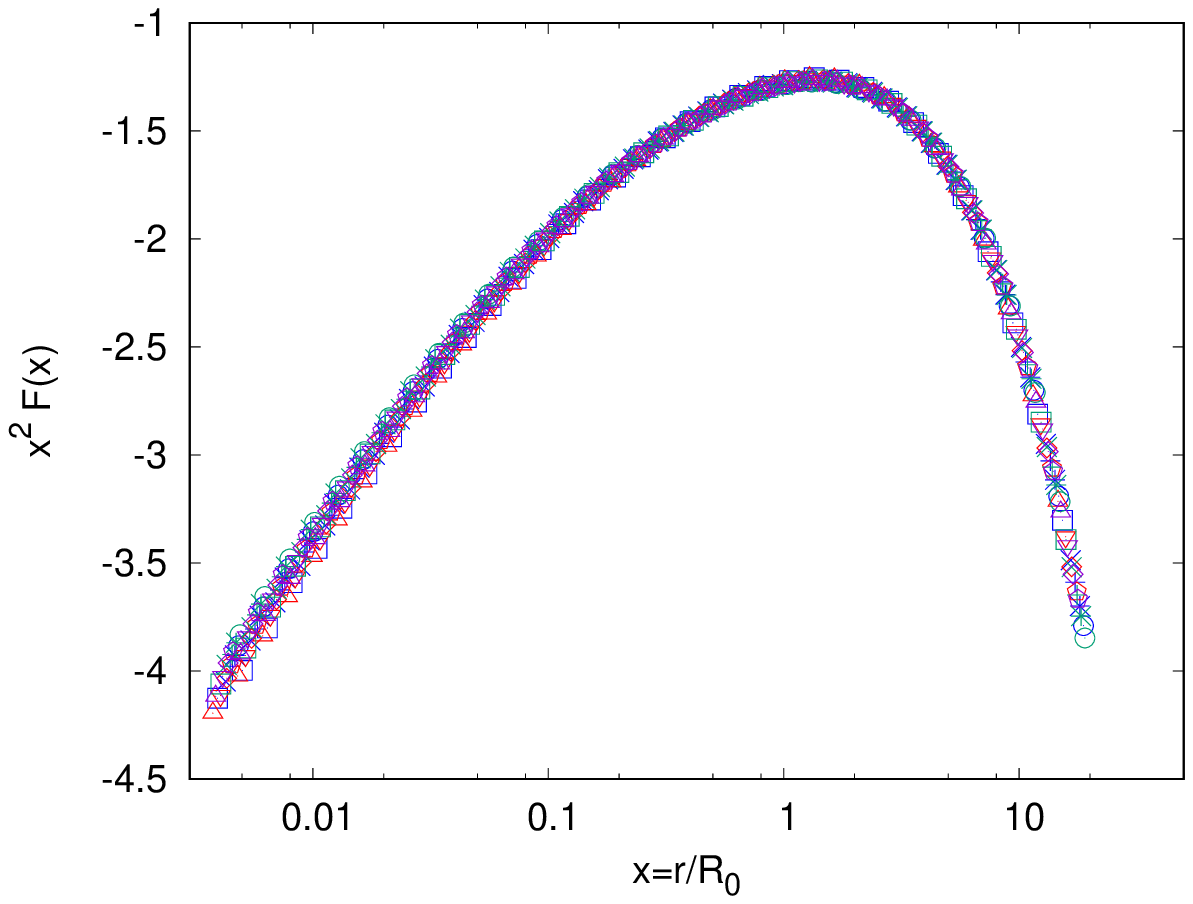}
\end{minipage}\hspace{2pc}%
\begin{minipage}{18pc}
{\footnotesize b)}\includegraphics[width=18pc]{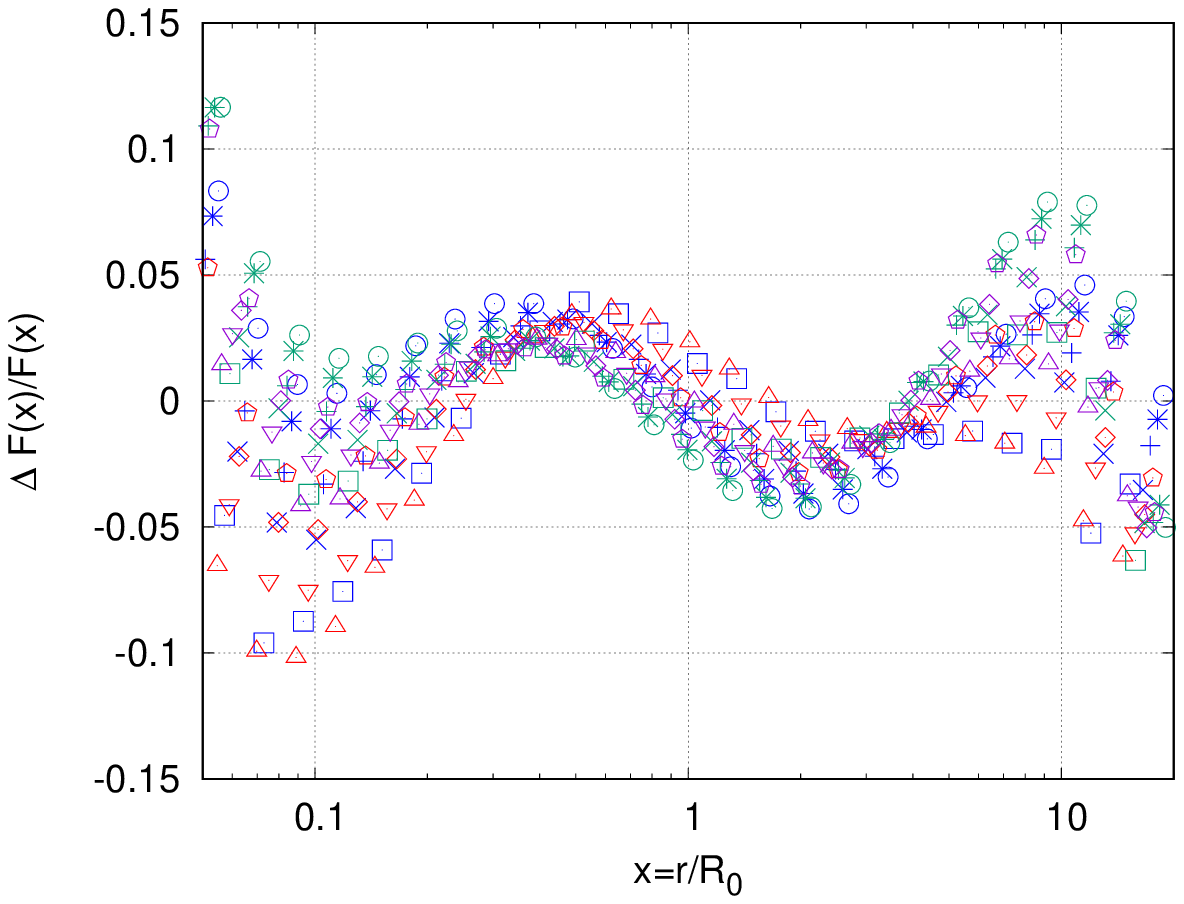}
\end{minipage} 
\caption{\label{f2}Invariant part (a) and relative uncertainties (b) of the scaling description of average lateral distribution of muons at sea level for vertical simulated EAS. All the parameters and designations are same as in Figure \ref{f1}.}
\end{figure}

\begin{figure}[t]
\begin{minipage}{18pc}
{\footnotesize a)}\includegraphics[width=18pc]{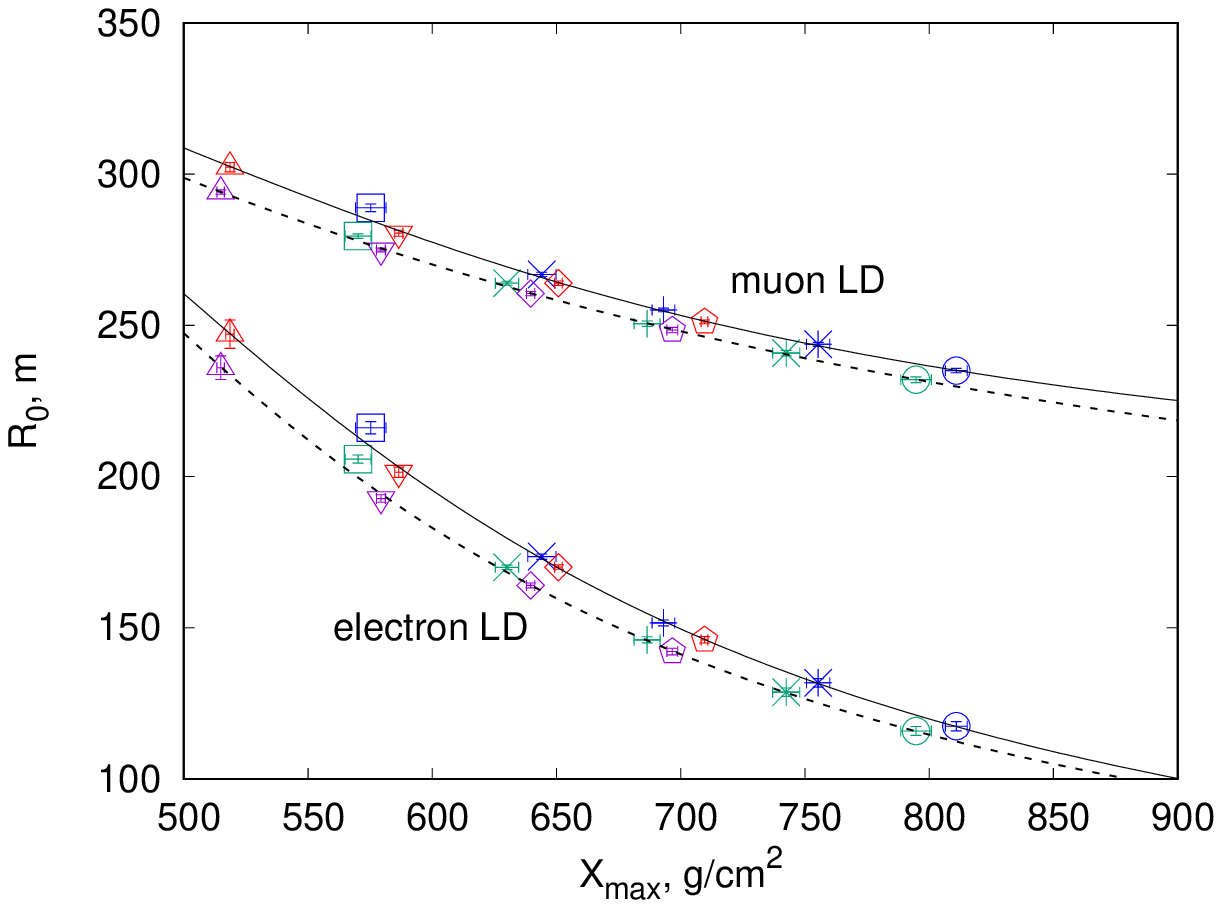}
\end{minipage}\hspace{2pc}%
\begin{minipage}{18pc}
	{\footnotesize b)}\includegraphics[width=18pc]{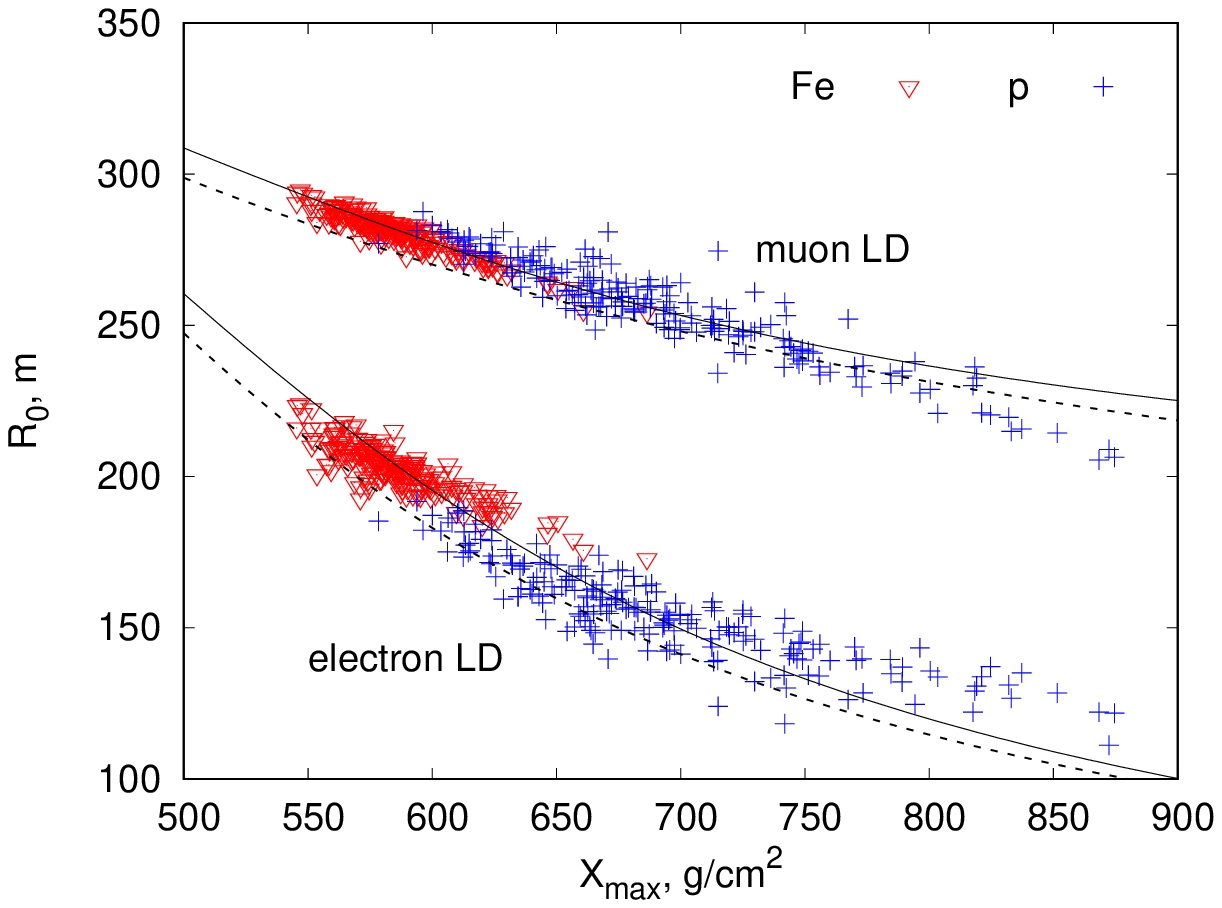}
\end{minipage} 
\caption{\label{f3}Scaling factors $R_0$ of lateral distributions of electrons (bottom) and muons (top) at sea level vs depth of shower maximum $X_{\rm max}$. a) Data points represent parameters of average vertical showers generated by protons and iron nuclei in the energy range of $\left(10^{15}-10^{19}\right)$~eV calculated using EPOS LHC and QGSJet hadronic interaction models. Designations are same as in Figure~\ref{f1}. $R_0(X_{\rm max})$ approximation curves for EPOS LHC (solid) and QGSJet (dotted) are shown to guide the eye. b) Parameters of individual showers generated by protons (blue crosses) and iron nuclei (red triangles) of $10^{17}$~eV simulated using EPOS LHC model are shown in a scatter plot. 200 events are included in each data set. The curves are average $R_0(X_{\rm max})$ approximations same as in Figure~\ref{f3}a).}
\end{figure}

In Figure~\ref{f1} the invariant part (a) and relative uncertainties (b) of the scaling description~(\ref{scaling1}) with polynomial $x^2F(x)$ fit of average lateral distribution of electrons with energy $>1$~MeV at sea level for vertical simulated EAS are shown. The invariant part of LD is given in more illustrative form $x^2F(x)$. In Figure~\ref{f2} the same results but for lateral distributions of muons with energy $>10$~MeV are presented.

It is seen from the figures, that lateral densities of electrons and muons, when considered separately, are well described by one-parametric scaling functions in the range of scaling variable $x=(0.05-20)$. The relative uncertainties do not exceed $15\%$ for electron LDF except iron induced showers with lowest energy $10^{16}$~eV, where the relative uncertainties are within $30\%$. For muon LD the accuracy of scaling descriptions is even better (mostly within 10\%). For charged particles distributions it was  found that scaling description with single $F(x)$ is violated considerably in radial distance range where electrons and muons give a comparable contribution to the local particle densities. The relative uncertainties reach $50\%$.

Examining the universality in relation between radial scale factors and depth of shower maximum we plot in Figure~\ref{f3}a) the $R_0(X_{\rm max})$ dependencies for electron (bottom) and muon (top) average LDs. Functional dependence $R_0(X_{\rm max})$ is observed for both electrons and muons. It is remarkable that the rates of change of scaling factors with increasing depth of maximum practically do not differ  for two hadronic interaction models.

In Figure~\ref{f3}b) the $R_0$ vs $X_{\rm max}$ scatter plot for individual showers induced by protons and iron nuclei of $10^{17}$~eV (EPOS LHC model) is demonstrated. 200 showers are included in each data set. Strong anticorrelation between the parameters allowing primary mass discrimination on the event-by-event basis should be noted.

\begin{figure}[t]
\begin{minipage}{18pc}
{\footnotesize a)}\includegraphics[width=18pc]{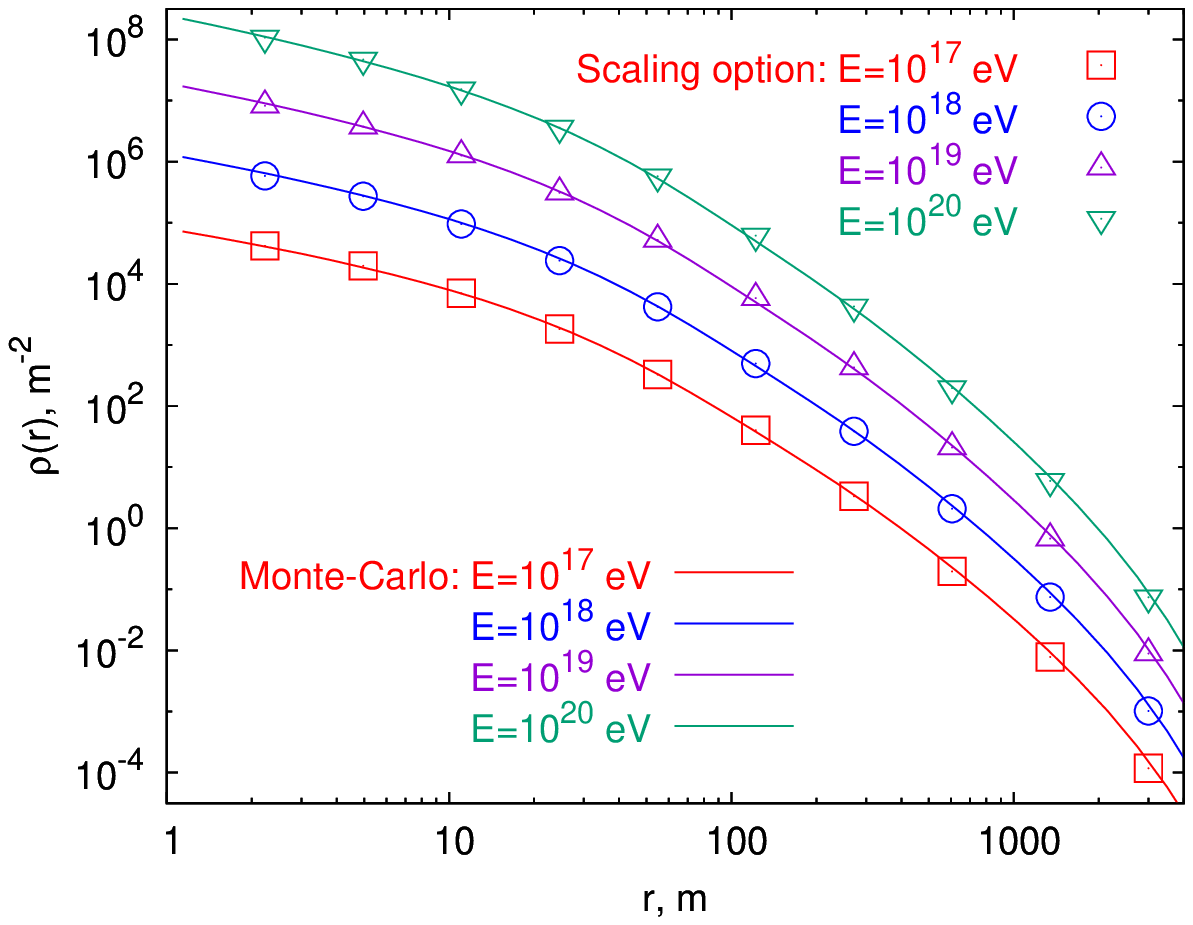}
\end{minipage}\hspace{2pc}%
\begin{minipage}{18pc}
{\footnotesize b)}\includegraphics[width=18pc]{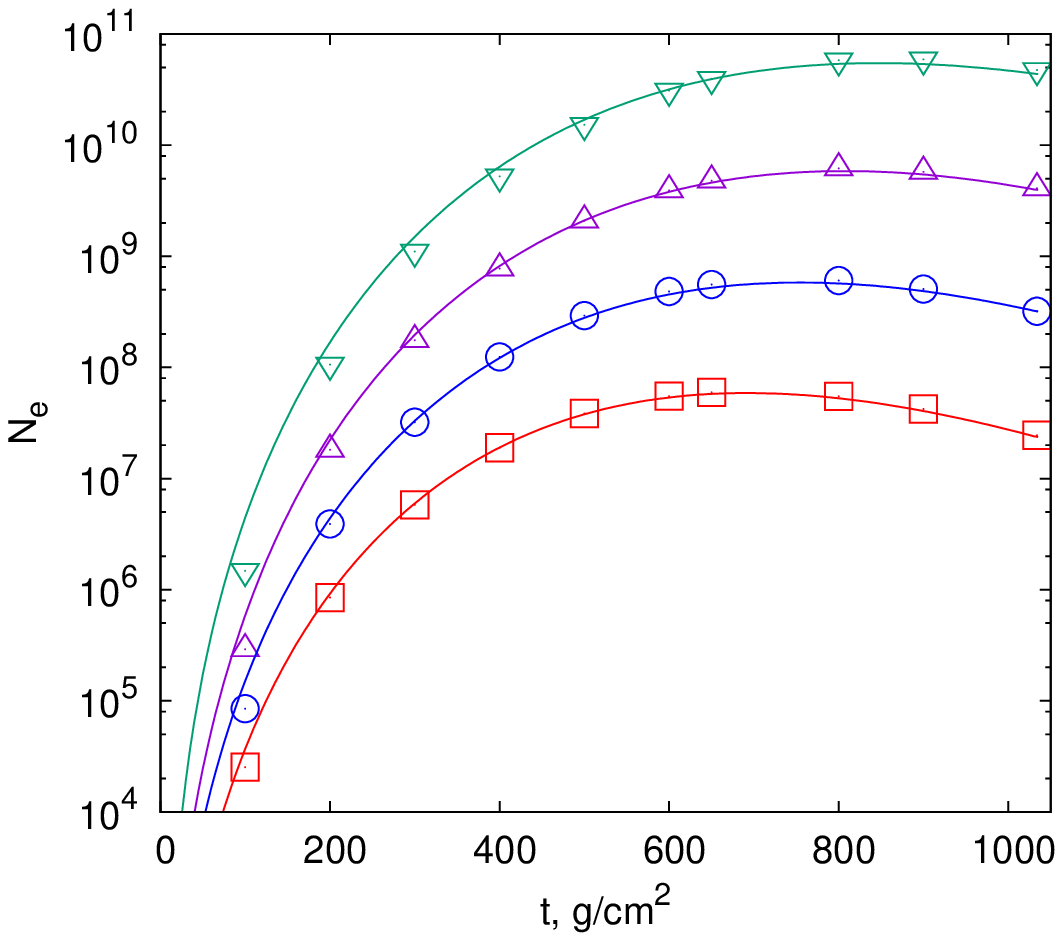}
\end{minipage} 
\caption{\label{MC_sc_comp}Comparison of Monte-Carlo simulation results (lines) with fast semi-analytical calculations based on the scaling formalism for lateral distributions (points). Primary protons of $\left(10^{17}-10^{20}\right)$~eV, EPOS LHC hadronic interaction model. Different symbols and colors correspond to different primary energies as described in the Figure. a) Average lateral distributions of electrons in the radial distance range $(2-3000)$~m. b) Average cascade curves.}
\end{figure}

Finally, it should be mentioned that due to its high accuracy for pure electromagnetic cascades in a wide range of radial distances, the scaling formalism can be successfully applied for fast semi-analytical EAS simulations. In  Figure~\ref{MC_sc_comp} the results of CORSIKA simulations with NKG-like option (scaling function with the appropriate values of $R_0$ is used instead of NKG to describe electromagnetic subshowers analytically) are shown in comparison with full Monte-Carlo treatment of EAS for primary protons of $\left(10^{17}-10^{20}\right)$~eV (EPOS LHC hadronic interaction model). Despite fluctuations suppression such a fast simulations could be very useful for the analysis of sensitivity to hadronic interaction models and meteorological effects.

\section{Summary}

Solving the mass composition problem is a key factor for understanding the origin of cosmic rays. A possible solution might be achieved with refined (multi-)hybrid measurements  together with generalizations of the analysis by revealing universal features, based on the intrinsic physical properties of air showers, evaluation of new parameters and functionals, which are weakly sensitive to the hadronic interaction model being good primary mass indicators.

In this paper we have demonstrated the extended scaling formalism, which enables accurate description of lateral distributions of electrons and muons by one-parametric scale-invariant functions in wide primary energy and radial distance ranges. The scale-invariance of LD and air shower universality manifesting through the functional dependence between radial scale factors and longitudinal shower age are both insensitive to hadronic interactions. The proposed analysis relies on general features of air showers and is thus less affected by sophisticated mixture of systematics in the measurement of the  cascade curve and lateral profile. Consequently, the proposed approach could be implemented for the present and future (multi-)hybrid air shower observations, especially in realizing the potential of Auger and Telescope Array observatories upgrades, as well as for re-analysis and cross-calibration of the data collected from different air shower arrays within the single method in a broad primary energy range.

\ack

This work was supported by Russian Foundation for Basic Research (grant \#16-02-01103~a)

\section*{References}
\bibliographystyle{iopart-num}
\bibliography{Raikin_ECRS_2018}

\end{document}